\documentclass[10pt]{article}
\usepackage{graphicx, changes}
\usepackage{booktabs}
\usepackage{subfig}
\usepackage{placeins}
\usepackage{tikz}
\usetikzlibrary{shapes, arrows.meta, positioning}
\usepackage{natbib, color}
\usepackage{verbatim,siunitx}
\usepackage{tgpagella}
\usepackage[para]{threeparttable}
\usepackage[T1]{fontenc}
\usepackage{ulem}
\usepackage{bbm}
\usepackage[hyphens]{url}
\usepackage{eurosym}
\usepackage{bbm}
\usepackage{float,comment}
\usepackage{amsfonts}
\usepackage{amsmath, amssymb, amsthm, graphicx}
\usepackage[onehalfspacing]{setspace}
\usepackage[a4paper,left=2.5cm, right=2.5cm,top=2.5cm, bottom=3cm]{geometry}
\usepackage{changepage}
\usepackage[nice]{nicefrac}

\newcommand{\eg}[1]{\citep[e.g.,][]{#1}}

\usepackage{multirow}

\providecommand{\U}[1]{\protect \rule{.1in}{.1in}}

\newtheorem{res}{Result}

\usepackage[multiple]{footmisc}

\begin{document}

\title{Public Support for Environmental Regulation: When Ideology Trumps Knowledge\thanks{The experiment was preregistered at AsPredicted, \protect\url{https://aspredicted.org/zqc8-z8y8.pdf}. This research and procedure complies with all relevant ethical regulations and was approved by the German Association for Experimental Economic Research (GfeW e.V., Institutional Review Board Certificate No. wPDjvfd2). This project has received funding from the German Research Foundation (DFG grant No. 490725816). The authors declare that they have no conflicts of interest related to the research presented in this paper. We thank Ben Mrakovcic and Maximilian Schmetzke for outstanding research assistance. We are grateful to Simon Cordes, Christoph Feldhaus, and Anna Ressi for valuable comments. This paper is dedicated to the memory of Reiner Hammer. All remaining errors are our own. 
}
}

\author{Markus Dertwinkel-Kalt \and Max R. P. Grossmann}

\bigskip
\date{\today}

\maketitle
\thispagestyle{empty}

\begin{abstract}
When environmental regulations are unpopular, policymakers often attribute resistance to information frictions and poor communication. We test this idea in the context of a major climate policy: Germany's Heating Law of 2023, which mandates the phase-out of fossil fuel heating. Through a survey experiment with property owners, we examine whether providing comprehensive information about the regulation's costs, requirements, and timeline affects adoption decisions and policy support. Despite successfully increasing factual knowledge, information provision has no significant effect on intended technology adoption, policy support, or incentivized measures of climate preferences. Instead, pre-existing environmental preferences and demographic characteristics emerge as the key predictors of responses to the regulation. A feeling that existing systems still work well and cost considerations dominate fossil fuel users' stated reasons for non-adoption, while independence from fossil fuels and perceived contributions to the common good drive adoption among switchers. Our findings suggest that opposition to climate policy stems from fundamental preference heterogeneity rather than information frictions. This has important implications for optimal policy design, highlighting potential limits of information provision in overcoming resistance to environmental regulation. The results also speak to broader questions in political economy about the relationship between knowledge, preferences, and support for policy reform.

\textit{JEL Classification}: D83, H31, Q48, Q58

\textit{Keywords}: Environmental regulation; technology adoption; information provision; political economy; climate policy

\end{abstract}

\clearpage
\setcounter{page}{1}

\setcounter{section}{0}

\section{Introduction}

Climate change represents one of the most significant economic challenges of our time. Measures to protect the environment, in turn, rely on public support. By amending the \textit{Gebäudeenergiegesetz} (“Building Energy Act”, henceforth: “Heating Law” or the Law) in September 2023 \citep{bmwk2023}, Germany established new rules that mandate heating with renewable energy and net-zero technologies for all buildings in the near future. The objective of the Law is to reduce carbon emissions in the building sector by achieving the climate targets set in the German \textit{Climate Protection Law} of \citeyear{bundesklimaschutzgesetz2019}. However, more than two-thirds of all households in 2023 still used fossil heating systems such as furnaces running on oil or gas \citep{bdew2023whd}, and a transition to heating systems compliant with the new Law is costly, with estimates reaching up to €1T---about one sixth of German GDP.\footnote{See \url{https://www.merkur.de/wirtschaft/verbot-oel-gas-heizungen-kosten-robert-habeck-energiewende-gesetz-92136808.html}.}

Very little is known about how to increase citizens' acceptance of net-zero measures.
We explore factual knowledge about such measures as a causal driver of attitudes.
The German Heating Law provides an ideal testing ground for this idea: the Law has been received poorly,\footnote{See, e.g., \url{https://www.zdf.de/phoenix/phoenix-nachgefragt/phoenix-heizungsgesetz-sehr-unbeliebt-100.html}.} and it has often been claimed that this is primarily due to poor communication.
Some media outlets have blamed politicians for poor communication,\footnote{See, e.g., \url{https://www.mdr.de/nachrichten/deutschland/politik/habeck-fehler-heizungsgesetz-kritik-waermepumpe-100.html}} some outlets blamed \emph{other} outlets for poor coverage of the Law,\footnote{See, e.g., \url{https://www.deutschlandfunk.de/heizungsgesetz-medien-heizhammer-bild-spiegel-100.html}.} and the Federal Minister for Economic Affairs and Climate Action, Robert Habeck, admitted his own communication failures in a TV interview, saying that he misestimated the public's mood after the winter following Russia's invasion of Ukraine.\footnote{See, e.g., \url{https://www.mdr.de/nachrichten/deutschland/politik/habeck-fehler-heizungsgesetz-kritik-waermepumpe-100.html}.}
Overall, it is believed that a major factor related to the unpopularity of the new Law is poor communication, and, as a consequence, voters being inadequately informed about the Law. Is it correct that the Law's unpopularity stems from insufficient or inadequate communication and information? Is knowledge causally related to attitudes---that is, can we increase citizens' knowledge and affect their view of the Law?

We run a large-scale survey experiment with German property owners to investigate how the support for controversial legislation depends on knowledge about that legislation. We investigate the role of knowledge on (i) their attitudes toward the Law, and (ii) their reactions to the Law in terms of their near-future heating system goals and their support for climate protection efforts. To do this, we leverage the changes made to the Law before its passage by experimentally varying truthful information provided about the Heating Law.

While selective presentation of information can easily produce predictable attitudinal shifts, our research question is: How does comprehensive, balanced knowledge affect outcomes? We specifically investigate the impact of exposing participants to a full measure of relevant information.
In a sense, our information provision experiment \citep{haaland2023designing,stantcheva2023run} can be understood as an investigation of the effect of political communication under a specific constraint: comprehensiveness and balance.
In one of our treatments (“Info Full”), we provide a short text explaining key features of the Law. Two further treatments represent strict subsets of Info Full; one of them, “Info Strict,” is thought to create an impression that the Law is strict, while “Info Lenient” was intended to create the opposite effect. Moreover, our study has a control group (baseline) of untreated respondents. After information is provided to subjects, we measure outcomes. We elicit motives for switching or not switching to net-zero heating technologies, and beliefs about these technologies and their adoption. Finally, respondents state their general preferences about government policy.

Our study reveals that neither property owners' attitudes toward the Law (i.e., its popularity), nor their planned responses to the Law (i.e., own heating choice), nor their incentivized attitudes towards climate protection are significantly affected by the information provided about the heating Law. This null effect is narrowly bounded, and a manipulation check confirms that our treatments are, by themselves, effective in increasing knowledge. Knowledge is also not correlated with outcomes in the untreated group. Overall, the Law is received poorly and knowledge about it is barely above chance. We thus provide evidence that attitudes and behaviors related to this legislation are not affected by better factual communication and the transmission of information. Rather, pre-existing ideology, in particular, a pro-environmental policy preference, as well as some some socio-demographic variables correlate significantly with attitudes towards the Law and behaviors. Those who stay with fossil technology are predominantly motivated by a feeling that the financial expenses necessary to switch are high and that their current system still works well, while switchers' motives are more nuanced. However, those who do not use fossil heating technology (or plan to switch away from fossil technology soon) appear to be motivated primarily by a desire for society to become independent of fossil imports and a feeling of pressure to switch.

By contributing to a literature on the acceptance of climate policies, our findings suggest that knowledge and facts may play a smaller role in shaping attitudes toward laws than ideology. While knowledge might still influence political values and policy support in the long run, our results indicate that ideological predispositions are a key determinant of attitudes toward specific regulations.

\paragraph{Related Literature.} We add to a literature on policies to phase out fossil fuel heating \citep{braungardt2023banning}. Public acceptance and support play critical roles in the implementation of climate policies \citep{fairbrother2022public, goerg2023public}. We also add to the literature on the role of communication and information strategies in increasing public support for carbon pricing \citep{woerner2023increase, cantner2023does}. The effectiveness of framing in enhancing public approval of climate policies has been shown by \citet{dasandi2022positive}. While we focus on a national law, international perspectives on fighting climate change and the attitudes toward climate policies are analyzed by \citet{dechezlepretre2022fighting}. Moreover, we demonstrate that objective knowledge about the Law is rather poor, similar to knowledge of other German energy policies \citep{dwk-pnas}. 

We add to a small literature in economics investigating the effects of the Heating Law. \cite{sollner2023neue} provides an analysis on the institutional--theoretical level. The author argues that the Law's measures will be ecologically ineffective, as they will not reduce Germany's overall $\text{CO}_2$ emissions due to existing emissions trading systems. Furthermore, the high cost of installing net-zero technologies such as heat pumps in existing buildings makes this an economically inefficient approach. \cite{sollner2023neue} claims that the \textit{dirigisme} inherent in the Law combined with substantial subsidies for homeowners is incompatible with the foundations of the German social market economy.

We also contribute to the literature on information campaigns, see \citet{haaland2023designing} for an overview and \citet{jager2024worker}, \citet{laudenbach2024beliefs}, and \citet{coibion2023does} for recent contributions. In particular, we add to the growing body of research on information campaigns related to economic policy \citep{allcott2015evaluating, coibion2022monetary, dwk-pnas}. Our focus is on a specific policy, namely, the \emph{Law}.

Finally, an emerging literature on misinformation has investigated how false news spread and perpetuate. Implicit in this literature is the idea that more factual information would lead to the enactment of policies subjectively preferred by scientists. We indirectly address this assumption, challenging the nexus between knowledge and support for policy. Similarly to \cite{barrera2020facts}, \cite{brandts2022dispelling} and \cite{dwk-pnas}, we find that factual knowledge can be increased, but that its effect on behaviors and attitudes is insignificant. Relatedly, \citet{d2021effective} find that policy communication is more effective at managing expectations when it emphasizes policy targets and objectives rather than the specific instruments used to achieve them, suggesting that explicitly communicating the policy itself may be of secondary importance.


\section{Research design}

\subsection{Background on the Law}

The amendment of the \textit{Gebäudeenergiegesetz} was enacted in September 2023 and aims to enhance energy efficiency and reduce carbon emissions in the building sector, supporting Germany's goal of achieving carbon neutrality by 2045. The Law has led to much public debate when it was first proposed in 2023 and many of the features that were initially discussed (such as a complete ban of oil and gas heating) were dropped. The final Law includes the extended operation of existing oil and gas heating systems beyond 2024, with a complete ban on their use coming into effect in 2045. The Law prohibits the installation of pure oil and gas heating systems from mid-2026 in large cities and from mid-2028 in smaller towns. In new buildings, the installation of such systems is prohibited immediately. From these dates, new heating systems must source at least 65\% of their energy from what the Law defines as “renewable sources.” Violations of these requirements can result in fines of up to €50,000. Additionally, new oil and gas heating systems installed before the respective bans can operate for up to 30 years but no longer than until 2045. Financial incentives are provided to property owners who install non-fossil heating systems, with subsidies covering at least 30\% of the installation costs, capped at €10,000.

The Law is expected to affect the heating market by increasing demand for renewable energy heating systems such as heat pumps, solar thermal systems, and biomass boilers. Financially, while the expenses are substantial, it is hoped that long-term savings on energy bills can compensate for initial investment costs for new heating systems. For tenants, a cost-sharing mechanism is included, whereby tenants only bear some part of the burden of a new heating system. Landlords can pass on at most 8\% of the investment costs annually to tenants. Economists have criticized that this regulation is inefficient because, unlike carbon pricing, it enforces carbon emission reductions in domains where they are particularly expensive to achieve \citep{sollner2023neue}.

\subsection{Basic survey design}

In February and March of 2024, we conducted a survey of German homeowners who either own their residences or act as landlords. We restricted the survey to homeowners, as they hold responsibility for heating technology decisions, whereas tenants generally lack this ability. For the survey, we use the panel of Bilendi \& respondi, which has 300,000 panelists in Germany. All adults
were eligible to participate in the study, provided they passed our screening question:
``Do you have the freedom to choose the type of heating in your home or in a rented property? This includes choosing between gas, oil, pellets, district heating, heat pumps, and other heating systems. Also click Yes if you are not the sole decision-maker, but have a say (e.g., as an owner of a condominium in a multi-family house). Also click No if you do not have heating. Also click No if you do not live in Germany.'' Bilendi \& respondi only provided us with completed surveys.

\begin{figure}
    \centering
    \resizebox{0.95\textwidth}{!}{%
    \begin{tikzpicture}[node distance=6cm, >=Stealth]
        \tikzstyle{block} = [rectangle, draw, fill=white, text centered, minimum height=2cm, minimum width=5cm]
        \tikzstyle{decision} = [rectangle, draw, fill=black!10, text centered, minimum height=1.5cm, minimum width=4cm]
        \tikzstyle{line} = [draw, ->]
    

        \node[block, align=center] (start) {\textbf{Started}\\$n=8{,}542$};
        \node[block, align=center, right =1cm of start] (process1) {\textbf{Proprietor or landlord}\\$n=3{,}460$};
        \node[decision, right=1cm of process1]  (decision1) {1. Demographics};
        
        \node[decision, below= 1cm of decision1] (decision2) {2. Information provision};
        \node[decision, left=1 cm of decision2] (decision3) {3. Attitudes};
        \node[decision, left=1cm of decision3] (decision4) {4. Quiz};
        \node[decision, left=1 cm of decision4, align=center] (decision5) {5. Beliefs, motives,\\dictator game};

        \node[decision, below =1cm of decision5] (decision6) {6. Policy preferences};
        \node[block, right of = decision6, align=center] (process2) {\textbf{Completed experiment}\\$n=2{,}995$};
        \node[block, right = 1cm of process2, align=center] (end) {\textbf{Passed attention check}\\$n=2{,}668$};

        \path[line] (start) -- (process1);
        \path[line] (process1) -- (decision1);
        \path[line] (decision1) -- (decision2);
        \path[line] (decision2) -- (decision3);
        \path[line] (decision3) -- (decision4);
        \path[line] (decision4) -- (decision5);
        \path[line] (decision5) -- (decision6);
        \path[line] (decision6) -- (process2);
        
        \path[line] (process2) -- (end);
    \end{tikzpicture}}

    \caption{Structure of the experiment}
    \label{structure}
\end{figure}
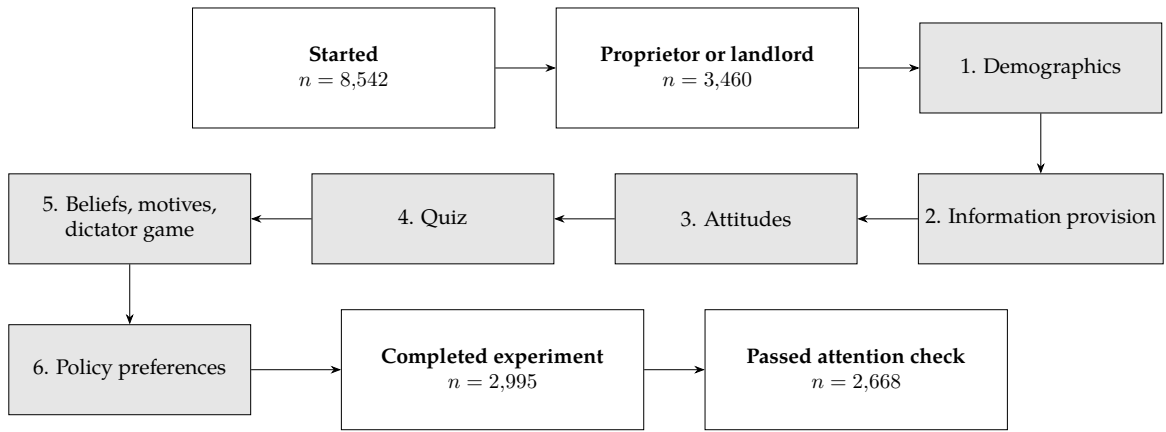

Figure \ref{structure} gives the structure of our survey. After a filter intended to restrict participation only to homeowners or landlords, subjects proceeded to a measurement of demographic characteristics. Subjects were subsequently exposed to one of our information treatments (see Section \ref{sec:infos}). Post-information, subjects gave their personal opinion of the Law and participated in a quiz about the Law. This quiz served to measure knowledge and as a manipulation check. Table \ref{tquestions} in the Appendix contains all questions from the quiz. Subjects were asked to give their beliefs about energy markets and heat pumps, and their motives for their choice of heating technology. Then, subjects stated beliefs about the motivations and choices of others and proceeded to a dictator game. Finally, respondents gave their preferences for government priorities. Subjects could receive bonus incentives only for the dictator game and their beliefs regarding the behavior of others. One participant (randomly selected from all subjects) had their choices in the dictator game implemented. Additionally, another participant (also randomly selected from all subjects) was chosen based on their performance in the belief elicitation task. All subjects received standard survey incentives for their participation (about €1 in total). The median respondent spent 12.5 minutes on our survey.

Our experiment provides knowledge to participants to study the effect of such knowledge on attitudes and behaviors regarding the Law. By experimentally inducing a change in knowledge, we can causally estimate knowledge's role in shaping public support for policy, free of confounds.

\subsection{Information treatments}
\label{sec:infos}

Our information provision experiment comprises four conditions (including the baseline). We exclusively provided truthful information. Here is the full content of our information treatments:

\vspace{1em}

\newcolumntype{P}[1]{>{\raggedright\arraybackslash}p{#1}}
\noindent\resizebox{\textwidth}{!}{%
\begingroup
\renewcommand{\arraystretch}{1.5} 

\begin{tabular}{@{}P{0.33\textwidth}P{0.33\textwidth}P{0.33\textwidth}@{}}
\small
\textit{Info Full} & \textit{Info Strict} & \textit{Info Lenient} \\
The new heating law allows the operation of oil and gas heaters beyond 2024, with a ban only coming into effect in 2045. &  & The new heating law allows the operation of oil and gas heaters beyond 2024, with a ban only coming into effect in 2045. \\
An immediate ban on oil and gas heaters was initially discussed, but it did not come into effect. &  & An immediate ban on oil and gas heaters was initially discussed, but it did not come into effect. \\
The installation of pure gas and oil heaters is banned in major cities from mid-2026, and in smaller cities from mid-2028. From then on, 65\% of the energy of a newly installed heating system must come from renewable sources. & The installation of pure gas and oil heaters is banned in major cities from mid-2026, and in smaller cities from mid-2028. From then on, 65\% of the energy of a newly installed heating system must come from renewable sources. &  \\
New oil and gas heaters may be operated for 30 years, but no later than 2045. Non-compliance can result in fines of up to €50,000. & New oil and gas heaters may be operated for 30 years, but no later than 2045. Non-compliance can result in fines of up to €50,000. &  \\
In new buildings, the installation of pure gas and oil heaters is banned effective immediately. &  &  \\
For the installation of a new heating system that operates without fossil fuels such as oil and gas, owners receive a subsidy of at least 30\%, but no more than €10,000. &  &  \\
Tenants contribute to the costs of a more modern heating system through their utility bills: Each year, up to 8\% of the investment costs can be passed on to tenants' utility bills. &  &  \\
\end{tabular}%
\endgroup%
}

\vspace{1em}

In the following, we describe the intentions behind the treatments. In all treatments, we provide information that is context-rich and balanced. Care was taken to present objective, fact-based information without advocating for particular positions. Recall that our objective is to investigate the effects of \emph{knowledge} on outcomes. Our setup does not include an ``active control'' treatment \citep{haaland2023designing} as we allow mere knowledge provision to influence outcomes. Moreover, a true active control is difficult to design. The information we provide is objective and cannot be easily inverted without deception, in contrast to projections about future macroeconomic patterns \eg{chopra2023home}.

\paragraph{Baseline.}

In our baseline condition, subjects received no information on the Law.

\paragraph{Info Full.}

In condition ``Info Full,'' respondents were exposed to a brief, pinpoint description of the central aspects of the Law. We used magazine and newspaper information targeted at homeowners to formulate this condition. Relative to the baseline, this treatment is the most informative. The German version of the information has 161 words.

\paragraph{Info Strict.}

Condition ``Info Strict'' emphasizes the Law's prohibition on the use of fossil technologies and the potential fines prescribed by the Law. We sought to create the impression that the Law is indeed quite strict. The German version has 57 words.

\paragraph{Info Lenient.}

Condition ``Info Lenient'' focuses on the Law allowing the use of fossil technologies until 2045 and that an immediate ban was initially proposed, but never made it into the final Law. Our intention with this condition was to invoke a feeling of the Law not being as strict as had originally been reported. The German version has 38 words.

\subsection{Measures}
\label{sec:measures}

We use four preregistered outcome variables in our study: (i) respondents' heating technology, (ii) their attitude towards the Law, (iii) their knowledge about the Law, and (iv) their donation to a climate charity. We now explain these outcome measures further.

First, all respondents stated their current heating system. If they planned to install a new one within the next year, their choice of technology for the new equipment was elicited. Proprietors responded for their own home; landlords for their rented property. Those who were both proprietors and landlords responded in only one randomly chosen role.\footnote{Below, we merge data from proprietors and landlords on technology usage.} For all respondents, we can code the utilization of fossil technology as follows:\footnote{Our classification uses the Law's classification: Gas and oil heating systems, as well as hybrid systems that use less than 65\% of renewable energy, are considered fossil. Heat pumps, district heating, hybrid systems with at least 65\% renewable energy sources and other heating technologies are considered non-fossil.} if a respondent presently uses fossil technology and does not plan to install a new system, they are coded as \textit{fossil}. Furthermore, if the respondent expects to install a new heating system and the target technology is fossil, they too are coded as users of fossil technology. All other respondents are coded as non-users of fossil technology. The resulting variable “Plan fossil usage” is binary. This variable is a forward-looking outcome variable. Below, we also report summary statistics on current usage; this variable is constructed similarly to “Plan fossil usage,” but since it concerns current usage, we call it “Use fossil now.”

Second, we elicited respondents' attitudes towards the Law. Our main focus will be on the agreement with the statement “The Law is sensible,” measured on a five-point Likert scale.

Third, each subject participates in a quiz comprising 12 questions on the new Law. All questions asked respondents to rate the correctness of a statement about the Law and its applications. Each question has a unique correct answer, some of which were provided in the information treatments (see Section \ref{sec:infos}). We calculate the “Knowledge Score,” a subject-specific count of correctly answered questions ranging from 0 to 12. Subjects were able to respond with “yes,” “no,” or “don’t know.” These questions were not incentivized to not give subjects an incentive to look up the correct answer on the internet.

Fourth, we gave each respondent €100 to split between three recipients: (i) themselves, (ii) a charity dedicated to promoting liberty and autonomy in Germany  (\textit{Prometheus – Das Freiheitsinstitut}), and (iii) a charity that supports effective climate protection (\textit{co2online – die gemeinnützige Energiespar-Beratung}). Both charities are non-profit entities and all contributions are tax-deductible. The former charity advocates for an open society and free markets. The latter charity, on which we focus in our analyses, provides information on climate measures to stakeholders and it offers non-profit pro-environmental consulting. Both of our charities are tax-exempt, public non-profit corporations, comparable to American 501(c)(3) status. These organizations seek to improve society through the communication of ideas while not being affiliated with political candidates or parties. We selected these charities because they reflect the primary publicly debated aspects regarding the Law: the restriction of individual freedom (charity (ii)) for the sake of climate protection (charity (iii)). By eliciting this donation decision, we aim to incentivize participants to reveal the relative importance they place on each of these aspects.

We elicited participants' motives to stay with a fossil or switch to a non-fossil heating technology. Homeowners and landlords were presented with either of two sets of motives: if they were fossil users, they were shown motives on not switching to non-fossil technologies, and vice versa for non-fossil users. The motives presented to fossil users related to the current heating working well, being more reliable, a switch of the heating system being too expensive or too effortful, feeling bullied by the Law or believing that the Law will not stay as it is. 
Motives presented to non-fossil users related to climate protection, independence of fossil imports, cost reasons, or having felt bullied into the new heating system. Motives for both fossil and non-fossil users also included an “Other” category. Respondents were allowed to put a custom motive into an open-ended text field. See Tables \ref{motives4fossil} and \ref{motives4nonfossil} in Section \ref{app:motives} in the Appendix for a list of all possible motives. We asked subjects to allocate 100\% of importance between these motives. For example, a subject may have allocated 80\% to “I want to contribute to climate protection,” 20\% to “I want us to become independent of fossil imports” and 0\% to all remaining motives. These additional measures allow us to ascertain the reasoning behind decisions.

In addition, we asked subjects to state their goals for general government policy. Goals we listed were economic growth, reduction of inequality, securing freedom of choice, preservation of the environment the climate and nature, preservation of traditions, and securing peace and security. We proceeded as with motives, with subjects dividing 100\% of importance among seven competing goals for policy. See Section \ref{app:policy} in the Appendix for the goals provided; again, we allowed the allocation to “Other” goals. This question informs us about the general political preferences of respondents. We here also define a new dummy variable, “Pro-Environmental Priority,” that equals one if the pro-environmental policy goal has maximal importance among all possible goals for each individual subject.\footnote{Note that our data on policy goals is necessarily compositional \citep{aitchison1982statistical} because the importance attached to goals is required to sum to 100\%. That compositional data create spurious negative correlations between the “columns” has been known since at least \cite{pearson1897mathematical}. Note that the concern about compositional data also applies to the data about motives; however, we analyze motives in a purely descriptive framework \citep{gerring2012mere}, while data on the pro-environmental priority will be used in regression analyses.}

We also elicited a wide range of demographic variables. Below, the omitted category for home type is ``apartment.''

\section{Results}

\subsection{Data and summary statistics}

\begin{table}[H]
\begin{center}
\begin{tabular}[t]{lccccc}
\toprule
Variable & Min. & Max. & Median & Mean & SD\\
\midrule
Knowledge Score & 0 & 12 & 7 & 6.593 & 2.261\\
Female & 0 & 1 & 0 & 0.431 & 0.495\\
Age & 18 & 75 & 46 & 46.794 & 12.814\\
High education & 0 & 1 & 1 & 0.604 & 0.489\\
High income & 0 & 1 & 0 & 0.461 & 0.499\\
Landlord & 0 & 1 & 0 & 0.296 & 0.457\\
Sole decider & 0 & 1 & 1 & 0.712 & 0.453\\

\addlinespace
\textit{Behaviors and attitudes}:\\
Plan fossil usage & 0 & 1 & 1 & 0.627 & 0.484\\
Use fossil now & 0 & 1 & 1 & 0.713 & 0.452\\
Law sensible & 1 & 5 & 3 & 2.715 & 1.356\\
Law good for climate & 1 & 5 & 3 & 2.919 & 1.355\\

\addlinespace
\textit{Dictator game}:\\
Give to self & 0 & 100 & 50 & 54.058 & 33.410\\
Give to pro-liberty charity & 0 & 100 & 20 & 21.296 & 21.695\\
Give to pro-climate charity & 0 & 100 & 25 & 24.645 & 24.193\\
\bottomrule
\end{tabular}

\caption{Summary statistics}
\label{demo}
\end{center}
\end{table}

Our final sample consists of 2,668 landlords and proprietors who completed the survey and passed a standard attention check \citep{stantcheva2023run}: the attention check required participants to select “agree” on a five-point Likert scale, which was placed among other scale items. Table \ref{demo} gives summary statistics about our sample's demographics, behaviors and attitudes. Particularly noteworthy is the knowledge score: a subject who randomized for each statement between “correct” and “not correct” would achieve a knowledge score of six. In the baseline, a knowledge score of only 6.18 is achieved. This implies that knowledge about the Law appears to be low (although it is significantly better than chance: $p = 0.03$ in a two-tailed $t$-test). A majority highly educated, having attained at least the German Abitur, i.e., graduated high school. About 46.1\% have high income, earning at least €4,000 per month after taxes. Almost 30\% are landlords, and more than 70\% can decide about heating technology without involving others.

While 71.3\% of respondents presently use fossil heating technology, only 62.7\% estimate that they will do so in one year. This means that the difference, about 8.6\% of respondents, plan to install a new non-fossil heating system within the next year. Only 32.6\% of those who plan to install any new system within the next year aim to install a fossil system.

We can now turn to the summary statistics of post-treatment variables. Almost 63\% of respondents use fossil technology for heating\footnote{This is below the German average, as 73\% of all homes in Germany are heated with fossil heating systems (see p.~51 in \cite{BDEW2024}). This gap can be explained by the fact that fewer older people participated in our study, who arguably tend to live in older homes. The average age in our study is 47 years, whereas the average homeowner in Germany is 58 years old (see \url{https://wohnglueck.de/artikel/immobilienbesitzer-deutschland-70720#e883cab1e266659c2c75f6e3e75d5993fd17ec73}).}; on our five-point Likert scale, the Law earns an average rating that is significantly below the midpoint of 3 ($p < 0.001$, two-tailed $t$-test), which captures the poor public reception of the Law:
31\% of respondents agree that the Law is sensible, while 43.8\% disagree. Among those who disagree, a majority strongly disagree. Additionally, 25.2\% position themselves at the midpoint. With respect to whether the Law helps the climate, opinions are evenly split: 37.7\% agree, and 37.9\% disagree. Notably, these two assessments are strongly correlated, as indicated by a Pearson correlation coefficient of $0.792$ ($p < 0.001$).

Finally, note the allocations in the dictator game. On average, respondents tend to allocate about half of the endowment to themselves, and split the remainder about equally between the pro-liberty and pro-climate charities, slightly favoring the latter.

\begin{table}
\begin{center}
\sisetup{parse-numbers=false, table-text-alignment=center}
\resizebox{\textwidth}{!}{%
\begin{tabular}{l S[table-format=4.6] S[table-format=4.6] S[table-format=4.6] S[table-format=4.6]}
\toprule
 & {Plan fossil usage} & {Law sensible} & {Give to climate org} & {Give to pro-liberty org} \\
\midrule
\emph{Intercept}               & 0.448^{***}  & 2.699^{***}  & 17.271^{***} & 22.019^{***} \\
                               & (0.053)      & (0.140)      & (2.603)      & (2.246)      \\
\emph{Treatment}: Info Full    & 0.013        & -0.042       & 0.173        & -1.004       \\
                               & (0.026)      & (0.070)      & (1.293)      & (1.202)      \\
\emph{Treatment}: Info Strict  & -0.002       & -0.070       & 1.228        & -1.143       \\
                               & (0.027)      & (0.073)      & (1.351)      & (1.199)      \\
\emph{Treatment}: Info Lenient & 0.010        & -0.028       & 0.418        & -1.432       \\
                               & (0.026)      & (0.072)      & (1.282)      & (1.184)      \\
Female                         & 0.033^{*}    & 0.023        & 1.449        & 0.723        \\
                               & (0.020)      & (0.054)      & (0.992)      & (0.901)      \\
Age                            & 0.004^{***}  & -0.008^{***} & 0.051        & -0.022       \\
                               & (0.001)      & (0.002)      & (0.038)      & (0.033)      \\
High education                 & -0.031       & 0.448^{***}  & 5.294^{***}  & -0.784       \\
                               & (0.020)      & (0.055)      & (1.009)      & (0.928)      \\
High income                    & -0.068^{***} & 0.107^{**}   & 1.074        & -0.219       \\
                               & (0.020)      & (0.054)      & (0.984)      & (0.891)      \\
Household size                 & -0.000       & 0.000        & 0.001        & -0.000       \\
                               & (0.000)      & (0.001)      & (0.002)      & (0.006)      \\
Landlord                       & -0.001       & 0.271^{***}  & 2.558^{**}   & 3.842^{***}  \\
                               & (0.022)      & (0.061)      & (1.091)      & (0.999)      \\
Duplex house                   & 0.052        & 0.025        & 2.103        & 2.967^{*}    \\
                               & (0.033)      & (0.092)      & (1.617)      & (1.544)      \\
Single-family house            & 0.024        & -0.048       & 0.615        & 0.108        \\
                               & (0.026)      & (0.072)      & (1.254)      & (1.123)      \\
Other home                     & 0.168        & -0.214       & 1.416        & 3.309        \\
                               & (0.124)      & (0.387)      & (5.520)      & (7.596)      \\
Townhouse                      & 0.127^{***}  & -0.068       & 2.970        & 1.710        \\
                               & (0.035)      & (0.100)      & (1.836)      & (1.563)      \\
Sole decider                   & 0.010        & 0.064        & -2.122^{*}   & -0.473       \\
                               & (0.022)      & (0.060)      & (1.138)      & (1.047)      \\
\midrule
R$^2$                          & 0.029        & 0.062        & 0.021        & 0.009        \\
Adj. R$^2$                     & 0.023        & 0.057        & 0.015        & 0.004        \\
Num. obs.                      &{2668}        &{2668}        &{2668}        &{2668}        \\
\bottomrule
\multicolumn{5}{l}{\scriptsize{$^{***}p<0.01$; $^{**}p<0.05$; $^{*}p<0.1$}}
\end{tabular}}
\caption{OLS regressions of behavior and attitudes on treatments and demographic variables\\{\footnotesize \emph{Plan fossil usage} is a binary variable that merges decisions from landlords and proprietors. \emph{Law sensible} is a Likert scale that ranges from 1 to 5. \emph{Give to climate org} is a euro amount (€0--€100) donated to the climate charity in the dictator game. Similarly, \emph{Give to pro-liberty org} is a euro amount (€0--€100) donated to the pro-liberty charity. HC3 standard errors in parentheses.}}
\label{regt1}
\end{center}
\end{table}

\subsection{Main results}

Table \ref{regt1} presents cross-sectional regressions of various outcome variables on covariates.\footnote{We deviate from our preregistration by excluding motives for heating decisions and beliefs about the heating law from these regressions. The reason for excluding motives is that questions differed between respondents who switched and those who did not. Beliefs about the heating law were not elicited shortly before conducting the survey due to space constraints. Additionally, while we do not directly include the Knowledge Score in our regressions, we provide further correlational evidence related to it below.} In all models, we use the treatments as a regressor. Based on recent findings on the robustness of “feelings integers” \citep{kaiser2022scientific}, we assume that Likert scales can enter our analyses as linear outcome variables. Similarly, in Table \ref{regt1k} in the Appendix we use the Knowledge Score as an explanatory variable. For all of our outcomes, in all specifications, knowledge about the Law shows no causal effect on behavior or attitudes. The same holds if we conduct a correlational analysis among those in the baseline treatment (Table \ref{regt1base} in the Appendix), that is, those who never received information. The standard errors attached to treatment and Knowledge Score coefficients are small across multiple specifications. Hence, these null effects are precisely estimated. Knowledge appears not to matter either as a cause nor as a mere correlational predictor of behaviors and attitudes. Figure \ref{fig:knowfine} in the Appendix further illustrates that there is no relationship between knowledge and the perception that the “Law is sensible.”

\begin{res}
	Knowledge about the Law is irrelevant for behaviors and attitudes. This holds both for the baseline and our treatments.
\end{res}

Table \ref{regknow} in the Appendix conducts regression analyses of the knowledge score on our treatments.
Column 2 highlights that older, more highly educated individuals and those in certain building types have a higher Knowledge Score, while females' Score is reduced. Our Full Info treatment led to an average increase of 1 correct response. This manipulation check demonstrates that our information treatments significantly increased knowledge.

\subsection{Political priorities predict behaviors and attitudes toward the law}

\begin{table}
\begin{center}
\sisetup{parse-numbers=false, table-text-alignment=center}
\resizebox{\textwidth}{!}{%
\begin{tabular}{l S[table-format=4.6] S[table-format=4.6] S[table-format=4.6] S[table-format=4.6]}
\toprule
 & {Plan fossil usage} & {Law sensible} & {Give to climate org} & {Give to pro-liberty org} \\
\midrule
\emph{Intercept}               & 0.476^{***}  & 2.446^{***} & 13.951^{***} & 22.950^{***} \\
                               & (0.053)      & (0.134)     & (2.549)      & (2.251)      \\
\emph{Treatment}: Info Full    & 0.008        & -0.001      & 0.722        & -1.158       \\
                               & (0.026)      & (0.067)     & (1.265)      & (1.199)      \\
\emph{Treatment}: Info Strict  & -0.004       & -0.058      & 1.382        & -1.186       \\
                               & (0.027)      & (0.070)     & (1.318)      & (1.194)      \\
\emph{Treatment}: Info Lenient & 0.009        & -0.025      & 0.447        & -1.440       \\
                               & (0.026)      & (0.069)     & (1.259)      & (1.180)      \\
Pro-Environmental Priority     & -0.096^{***} & 0.873^{***} & 11.462^{***} & -3.212^{***} \\
                               & (0.020)      & (0.051)     & (1.020)      & (0.833)      \\
\midrule
Demographic controls           & {Yes}        & {Yes}       & {Yes}        & {Yes}        \\
R$^2$                          & 0.037        & 0.152       & 0.069        & 0.014        \\
Adj. R$^2$                     & 0.032        & 0.147       & 0.064        & 0.008        \\
Num. obs.                      &{2668}        &{2668}       &{2668}        &{2668}        \\
\bottomrule
\multicolumn{5}{l}{\scriptsize{$^{***}p<0.01$; $^{**}p<0.05$; $^{*}p<0.1$}}
\end{tabular}}
\caption{OLS regressions of behavior and attitudes on the pro-environmental priority\\{\footnotesize This Table is identical to Table \ref{regt1}, but with “Pro-Environmental Priority” as an additional regressor. \emph{Plan fossil usage} is a binary variable that merges decisions from landlords and proprietors. \emph{Law sensible} is a Likert scale that ranges from 1 to 5. \emph{Give to climate org} is a euro amount (€0--€100) donated to the climate charity in the dictator game. Similarly, \emph{Give to pro-liberty org} is a euro amount (€0--€100) donated to the pro-liberty charity. Demographic controls: Female, age, high education, high income, household size, landlord, duplex house, single-family house, other home, townhouse, sole decider. HC3 standard errors in parentheses.}}
\label{regt2a}
\end{center}
\end{table}

Given that knowledge is not important for behaviors and attitudes regarding the Law, what has better explanatory power? In this subsection, we explore whether fundamental preferences toward the goals of general policy have predictive power. To elicit these, we asked subjects to allocate 100\% of importance among seven competing goals for general government policy, and we derive an indicator variable, “Pro-Environmental Priority,” to denote that a subject ascribes maximum importance to pro-environmental policy (see Section \ref{sec:measures} and Section
 \ref{app:policy} in the Appendix for details).

Table \ref{regt2a} presents regressions of behavior and attitudes on treatment and “Pro-Environmental Priority.”\footnote{Table \ref{regt2} in the Appendix presents an alternative specification that includes all policy goals (with the “Other” policy goal left out because of multicollinearity).} Table \ref{regt2a} is otherwise identical to Table \ref{regt1}. The results indicate that having the pro-environmental priority is strongly predictive of behaviors \emph{and} attitudes. However, there is no significant relation to knowledge about the Law. Column 4 of Table \ref{regt2a} indicates that donations to the climate charity are also predicted by the pro-environmental priority, indicating a strong consistency between these items. Figure \ref{fig:priofine} in the Appendix demonstrates a striking difference in the distributions of “Law sensible” that occurs between those with the pro-environmental priority and those without.

Furthermore, Table \ref{regt1} shows that demographic variables show a significant relationship to attitudes and behavior, even though knowledge on its own does not. For example, highly educated individuals tend to have increased support for the Law and increased donations to a climate organization. On the other hand, they have no more knowledge and show no more use of net-zero technology. Those with higher income, however, appear to have more knowledge and use for non-fossil heating, but they show no effect on support for the Law or the climate organization.

\begin{res}
Behaviors and attitudes are significantly predicted by demographics and a pro-environmental policy priority.
\end{res}

Figure \ref{fig:combined_motives} in the Appendix shows average motives between fossil and non-fossil users as well as average beliefs about the behaviors of others. Table \ref{mot1tab} in the Appendix shows full summary statistics. For fossil users, that the current system still works well and that substantial financial expenses would be necessary to switch appear to be the major motives for staying with fossil technology. For non-fossil users, the picture is more mixed, with all motives attaining substantial allocations, even the “Other” motive. However, the most prominent motive appears to be a willingness to become independent of fossil imports, followed by a feeling of being pressured into using net-zero technologies and a desire to contribute to the common good of protecting the climate. Pressure to use non-fossil technology is thought to be the major motive in the decision-making of others.
Regarding users of fossil technology, they tend to overestimate financial motives in others, and to underestimate reliability concerns. Both of these differences are highly significant (paired $t$-test, maximum $p < 0.001$).

\begin{table}
\begin{center}
\sisetup{parse-numbers=false, table-text-alignment=center}
\begin{tabular}{l S[table-format=4.6] S[table-format=4.6]}
\toprule
 & {Law good for climate} & {Belief: percent non-fossil} \\
\midrule
\emph{Intercept}               & 2.551^{***} & 38.193^{***} \\
                               & (0.138)     & (2.448)      \\
\emph{Treatment}: Info Full    & -0.032      & -1.739^{*}   \\
                               & (0.067)     & (1.022)      \\
\emph{Treatment}: Info Strict  & -0.079      & -1.416       \\
                               & (0.071)     & (1.035)      \\
\emph{Treatment}: Info Lenient & -0.085      & 0.137        \\
                               & (0.070)     & (1.057)      \\
Pro-Environmental Priority     & 0.857^{***} & 2.942^{***}  \\
                               & (0.052)     & (0.782)      \\
Belief is about proprietors    &             & -0.111       \\
                               &             & (1.548)      \\
\midrule
Demographic controls           & {Yes}       & {Yes}        \\
R$^2$                          & 0.131       & 0.078        \\
Adj. R$^2$                     & 0.126       & 0.072        \\
Num. obs.                      &{2668}       &{2668}        \\
\bottomrule
\multicolumn{3}{l}{\scriptsize{$^{***}p<0.01$; $^{**}p<0.05$; $^{*}p<0.1$}}
\end{tabular}
\caption{OLS regressions of assessment of the Law's climate effects and beliefs on the pro-environmental priority and own behavior\\{\footnotesize \emph{Law good for climate} is a Likert scale that ranges from 1 to 5. \emph{Belief: percent non-fossil} is a percentage (0--100) about the proportion thought to have switched to non-fossil heating or planning such a switch in the next year. Demographic controls: Female, age, high education, high income, household size, landlord, duplex house, single-family house, other home, townhouse, sole decider. HC3 standard errors in parentheses.}}
\label{regt3}
\end{center}
\end{table}

We also find evidence in line with several well-known established biases. Column 1 in Table \ref{regt3} shows a cross-sectional regression of the assessment of whether the Law is seen as good for the climate on our treatments, the pro-environmental priority and own behavior. The pro-environmental priority is highly correlated with a more positive assessment of the Law's climate impact, while the use of fossil technology is associated with a more negative assessment. This can be viewed as an expression of “motivated reasoning,” the idea that people hold systematically different beliefs to view their own behavior more favorably \eg{amelio2023motivated,zimmermann2020dynamics,kunda1990case}.
In Column 2 in Table \ref{regt3}, we investigate false consensus bias---the tendency to overestimate how much others share the own beliefs, attitudes, or behaviors. \eg{ross1977false}. First, note that landlords' and homeowners' own behaviors are strongly correlated with the expected behavior of others. Those who use fossil heating technology believe that fewer others have switched to non-fossil technologies. Column 2 also indicates that those with the pro-environmental priority are optimistic regarding this proportion. This suggests that also false consensus bias plays a role in our context.

\subsection{The effect of fine-grained knowledge}

We can regress our outcome variables on responses to individual questions. Table \ref{regt9} in the Appendix reveals that knowing the answer to any particular question still has a low, if any, relationship to the outcomes.

The coefficients reveal opposing signs across different variables. While acknowledging the limitations of correlational analysis and multiple hypothesis testing, we observe that detailed policy knowledge influences behavior and attitudes in complex ways. For example, awareness of the Law's penalties (Q6) is associated with reduced fossil fuel use but also lower approval of the Law itself. In contrast, knowledge of subsidies (Q8) correlates with both decreased fossil fuel use and increased support for the Law, though these relationships are modest. These findings inform our core research question: how the provision of balanced information impacts outcomes. It is plausible that positively valenced and negatively valenced pieces of information “cancel out” when information is required to be contextual and comprehensive.

By contrast, subjects' environmental priorities consistently exhibit a strong correlation. From this additional robustness check, we conclude that ideology indeed outweighs knowledge—both “knowledge writ large” and fine-grained, issue-specific knowledge.

\subsection{Limitations}

In this subsection we briefly discuss potential limitations of our study. First, the sample in this study is not representative of the broader German population. Our focus lies specifically on households directly affected by the German Heating Law, that is, those that choose their heating system themselves, especially home proprietors and landlords. A representative profile of German homeowners and landlords is not available (including from sources such as the \textit{German Census} or \textit{SOEP}), so it is impossible for us to recruit a sample that is representative of this specific group. We have however no good reason to assume that our sample of self-selected participants is biased in a way that systematically distorts any of our findings. Since homeowners and landlords are the main population responsible for choosing between heating technologies, our results can be viewed as an upper bound on the effect of knowledge on support for policy among those able to act.

Second, our main outcome variables---attitudes and intended reactions to the law---are not incentivized. We do not see, however, how this could plausibly affect our findings. To account for the fact that our intervention cannot cause the adoption of non-fossil heating technologies immediately, that is, during our study, we asked for participants' intended heating technology choice for the next year.

Third, one could ask whether experimenter demand effects could affect our findings. \cite{haaland2023designing} discuss demand effects in information provision experiments (pp. 21ff.).
Our implicit model of information provision provides little reason to suppose that our treatments affect outcomes by themselves; rather, they are thought to influence knowledge, which in turn causes the outcomes. Indeed, the robust and narrowly bounded null effects in Table \ref{regt1} on the information treatments show no presence of direct effects. Such direct effects could be viewed as demand effects. Thus, we also add to an emerging literature suggesting that demand effects may not be relevant concerns for experimental research \citep{de2018measuring}. Moreover, Table \ref{regtgoaltreat} in the Appendix indicates that policy preferences were not influenced by our treatment, again hinting at the absence of demand effects.

\section{Conclusion}

Our findings underscore the limited influence of factual knowledge on public support for climate policy, challenging the common assertion by policymakers that unpopularity stems from poor communication and poor information dissemination.  Our results suggest that support for climate policy is more strongly anchored in pre-existing ideology, in particular, a pro-environmental policy preference, than in understanding of policy details.

\bibliographystyle{ecca}
\bibliography{lit-heating}

\begin{thebibliography}{33}
\providecommand{\natexlab}[1]{#1}

\bibitem[{Aitchison(1982)}]{aitchison1982statistical}
\textsc{Aitchison, J.} (1982). The statistical analysis of compositional data.
  \textit{Journal of the Royal Statistical Society: Series B}, \textbf{44}~(2),
  139--160.

\bibitem[{Allcott and Taubinsky(2015)}]{allcott2015evaluating}
\textsc{Allcott, H.} and \textsc{Taubinsky, D.} (2015). Evaluating behaviorally
  motivated policy: Experimental evidence from the lightbulb market.
  \textit{American Economic Review}, \textbf{105}~(8), 2501--2538.

\bibitem[{Amelio and Zimmermann(2023)}]{amelio2023motivated}
\textsc{Amelio, A.} and \textsc{Zimmermann, F.} (2023). Motivated memory in
  economics—a review. \textit{Games}, \textbf{14}~(1), 15.

\bibitem[{Barrera \textit{et~al.}(2020)Barrera, Guriev, Henry and
  Zhuravskaya}]{barrera2020facts}
\textsc{Barrera, O.}, \textsc{Guriev, S.}, \textsc{Henry, E.} and
  \textsc{Zhuravskaya, E.} (2020). Facts, alternative facts, and fact checking
  in times of post-truth politics. \textit{Journal of Public Economics},
  \textbf{182}, 104123.

\bibitem[{Brandts \textit{et~al.}(2022)Brandts, Busom, Lopez-Mayan and
  Panad{\'e}s}]{brandts2022dispelling}
\textsc{Brandts, J.}, \textsc{Busom, I.}, \textsc{Lopez-Mayan, C.} and
  \textsc{Panad{\'e}s, J.} (2022). Dispelling misconceptions about economics.
  \textit{Journal of Economic Psychology}, \textbf{88}, 102461.

\bibitem[{Braungardt \textit{et~al.}(2023)Braungardt, Tezak, Rosenow and
  B{\"u}rger}]{braungardt2023banning}
\textsc{Braungardt, S.}, \textsc{Tezak, B.}, \textsc{Rosenow, J.} and
  \textsc{B{\"u}rger, V.} (2023). Banning boilers: An analysis of existing
  regulations to phase out fossil fuel heating in the {EU}. \textit{Renewable
  and Sustainable Energy Reviews}, \textbf{183}, 113442.

\bibitem[{{Bundes-Klimaschutzgesetz}(2019)}]{bundesklimaschutzgesetz2019}
\textsc{{Bundes-Klimaschutzgesetz}} (2019). {Bundes-Klimaschutzgesetz}.
  Available at \url{https://www.gesetze-im-internet.de/ksg/BJNR251310019.html}.

\bibitem[{{Bundesverband der Energie- und
  Wasserwirtschaft}(2023)}]{bdew2023whd}
\textsc{{Bundesverband der Energie- und Wasserwirtschaft}} (2023). {Wie heizt
  Deutschland 2023?} Available at
  \url{https://www.bdew.de/media/documents/231221-BDEW-WHD2023.pdf}.

\bibitem[{{Bundesverband der Energie- und Wasserwirtschaft
  (BDEW)}(2024)}]{BDEW2024}
\textsc{{Bundesverband der Energie- und Wasserwirtschaft (BDEW)}} (2024). Die
  energieversorgung 2024 -- jahresbericht des bdew.
  \url{https://www.bdew.de/media/documents/2024_12_18_Die_Energieversorgung_2024_Final.pdf},
  accessed: 2025-03-12.

\bibitem[{Cantner and Rolvering(2023)}]{cantner2023does}
\textsc{Cantner, F.} and \textsc{Rolvering, G.} (2023). Does information help
  to overcome resistance to carbon pricing? {E}vidence from a survey
  experiment. Available at
  \url{https://papers.ssrn.com/sol3/papers.cfm?abstract_id=4704632}.

\bibitem[{Chopra \textit{et~al.}(2023)Chopra, Roth and
  Wohlfart}]{chopra2023home}
\textsc{Chopra, F.}, \textsc{Roth, C.} and \textsc{Wohlfart, J.} (2023). Home
  price expectations and spending: Evidence from a field experiment.
  \textit{Available at SSRN 4452588}.

\bibitem[{Coibion \textit{et~al.}(2023)Coibion, Georgarakos, Gorodnichenko and
  Van~Rooij}]{coibion2023does}
\textsc{Coibion, O.}, \textsc{Georgarakos, D.}, \textsc{Gorodnichenko, Y.} and
  \textsc{Van~Rooij, M.} (2023). How does consumption respond to news about
  inflation? field evidence from a randomized control trial. \textit{American
  Economic Journal: Macroeconomics}, \textbf{15}~(3), 109--152.

\bibitem[{Coibion \textit{et~al.}(2022)Coibion, Gorodnichenko and
  Weber}]{coibion2022monetary}
\textsc{---}, \textsc{Gorodnichenko, Y.} and \textsc{Weber, M.} (2022).
  Monetary policy communications and their effects on household inflation
  expectations. \textit{Journal of Political Economy}, \textbf{130}~(6),
  1537--1584.

\bibitem[{D'Acunto \textit{et~al.}(2021)D'Acunto, Hoang, Paloviita and
  Weber}]{d2021effective}
\textsc{D'Acunto, F.}, \textsc{Hoang, D.}, \textsc{Paloviita, M.} and
  \textsc{Weber, M.} (2021). \textit{Effective policy communication: Targets
  versus instruments}. Tech. rep., KIT Working Paper Series in Economics.

\bibitem[{Dasandi \textit{et~al.}(2022)Dasandi, Graham, Hudson, Jankin, van
  Heerde-Hudson and Watts}]{dasandi2022positive}
\textsc{Dasandi, N.}, \textsc{Graham, H.}, \textsc{Hudson, D.}, \textsc{Jankin,
  S.}, \textsc{van Heerde-Hudson, J.} and \textsc{Watts, N.} (2022). Positive,
  global, and health or environment framing bolsters public support for climate
  policies. \textit{Communications Earth \& Environment}, \textbf{3}~(1), 239.

\bibitem[{De~Quidt \textit{et~al.}(2018)De~Quidt, Haushofer and
  Roth}]{de2018measuring}
\textsc{De~Quidt, J.}, \textsc{Haushofer, J.} and \textsc{Roth, C.} (2018).
  Measuring and bounding experimenter demand. \textit{American Economic
  Review}, \textbf{108}~(11), 3266--3302.

\bibitem[{Dechezlepr{\^e}tre \textit{et~al.}(2023)Dechezlepr{\^e}tre, Fabre,
  Kruse, Planterose, Chico and Stantcheva}]{dechezlepretre2022fighting}
\textsc{Dechezlepr{\^e}tre, A.}, \textsc{Fabre, A.}, \textsc{Kruse, T.},
  \textsc{Planterose, B.}, \textsc{Chico, A.~S.} and \textsc{Stantcheva, S.}
  (2023). Fighting climate change: International attitudes toward climate
  policies. Available at
  \url{https://socialeconomicslab.org/wp-content/uploads/2024/05/international_attitudes_toward_climate_change.pdf}.

\bibitem[{Dertwinkel-Kalt \textit{et~al.}(2024)Dertwinkel-Kalt, Feldhaus,
  Ockenfels and Sutter}]{dwk-pnas}
\textsc{Dertwinkel-Kalt, M.}, \textsc{Feldhaus, C.}, \textsc{Ockenfels, A.} and
  \textsc{Sutter, M.} (2024). Household reduction of gas consumption in the
  energy crisis is not explained by individual economic incentives.
  \textit{Proceedings of the National Academy of Sciences}, \textbf{121}~(48),
  e2411740121.

\bibitem[{Fairbrother(2022)}]{fairbrother2022public}
\textsc{Fairbrother, M.} (2022). Public opinion about climate policies: A
  review and call for more studies of what people want. \textit{PLoS Climate},
  \textbf{1}~(5), e0000030.

\bibitem[{GEG(2023)}]{bmwk2023}
\textsc{GEG} (2023). {Gesetz zur Einsparung von Energie und zur Nutzung
  erneuerbarer Energien zur Wärme- und Kälteerzeugung in Gebäuden
  (Gebäudeenergiegesetz - GEG)}. Available at
  \url{https://www.gesetze-im-internet.de/geg/GEG.pdf}.

\bibitem[{Gerring(2012)}]{gerring2012mere}
\textsc{Gerring, J.} (2012). Mere description. \textit{British Journal of
  Political Science}, \textbf{42}~(4), 721--746.

\bibitem[{Goerg \textit{et~al.}(2023)Goerg, Pondorfer and
  St{\"o}hr}]{goerg2023public}
\textsc{Goerg, S.}, \textsc{Pondorfer, A.} and \textsc{St{\"o}hr, V.} (2023).
  Public support for more ambitious climate policies. Available at
  \url{https://ideas.repec.org/p/aiw/wpaper/30.html}.

\bibitem[{Haaland \textit{et~al.}(2023)Haaland, Roth and
  Wohlfart}]{haaland2023designing}
\textsc{Haaland, I.}, \textsc{Roth, C.} and \textsc{Wohlfart, J.} (2023).
  Designing information provision experiments. \textit{Journal of Economic
  Literature}, \textbf{61}~(1), 3--40.

\bibitem[{J{\"a}ger \textit{et~al.}(2024)J{\"a}ger, Roth, Roussille and
  Schoefer}]{jager2024worker}
\textsc{J{\"a}ger, S.}, \textsc{Roth, C.}, \textsc{Roussille, N.} and
  \textsc{Schoefer, B.} (2024). Worker beliefs about outside options.
  \textit{The Quarterly Journal of Economics}, \textbf{139}~(3), 1505--1556.

\bibitem[{Kaiser and Oswald(2022)}]{kaiser2022scientific}
\textsc{Kaiser, C.} and \textsc{Oswald, A.~J.} (2022). The scientific value of
  numerical measures of human feelings. \textit{Proceedings of the National
  Academy of Sciences}, \textbf{119}~(42), e2210412119.

\bibitem[{Kunda(1990)}]{kunda1990case}
\textsc{Kunda, Z.} (1990). The case for motivated reasoning.
  \textit{Psychological bulletin}, \textbf{108}~(3), 480.

\bibitem[{Laudenbach \textit{et~al.}(2024)Laudenbach, Weber, Weber and
  Wohlfart}]{laudenbach2024beliefs}
\textsc{Laudenbach, C.}, \textsc{Weber, A.}, \textsc{Weber, R.} and
  \textsc{Wohlfart, J.} (2024). Beliefs about the stock market and investment
  choices: Evidence from a survey and a field experiment. \textit{The Review of
  Financial Studies}, p. hhae063.

\bibitem[{Pearson(1897)}]{pearson1897mathematical}
\textsc{Pearson, K.} (1897). Mathematical contributions to the theory of
  evolution—on a form of spurious correlation which may arise when indices
  are used in the measurement of organs. \textit{Proceedings of the Royal
  Society of London}, \textbf{60}~(359-367), 489--498.

\bibitem[{Ross \textit{et~al.}(1977)Ross, Greene and House}]{ross1977false}
\textsc{Ross, L.}, \textsc{Greene, D.} and \textsc{House, P.} (1977). The
  “false consensus effect”: An egocentric bias in social perception and
  attribution processes. \textit{Journal of Experimental Social Psychology},
  \textbf{13}~(3), 279--301.

\bibitem[{S{\"o}llner(2023)}]{sollner2023neue}
\textsc{S{\"o}llner, F.} (2023). Das neue {G}eb{\"a}udeenergiegesetz.
  \textit{Wirtschaftsdienst}, \textbf{103}~(9), 619--623.

\bibitem[{Stantcheva(2023)}]{stantcheva2023run}
\textsc{Stantcheva, S.} (2023). How to run surveys: A guide to creating your
  own identifying variation and revealing the invisible. \textit{Annual Review
  of Economics}, \textbf{15}, 205--234.

\bibitem[{Woerner \textit{et~al.}(2023)Woerner, Imai, Pace and
  Schmidt}]{woerner2023increase}
\textsc{Woerner, A.}, \textsc{Imai, T.}, \textsc{Pace, D.} and \textsc{Schmidt,
  K.} (2023). How to increase public support for carbon pricing. Available at
  \url{https://www.econstor.eu/handle/10419/282180}.

\bibitem[{Zimmermann(2020)}]{zimmermann2020dynamics}
\textsc{Zimmermann, F.} (2020). The dynamics of motivated beliefs.
  \textit{American Economic Review}, \textbf{110}~(2), 337--363.

\end{thebibliography}
\section*{Appendix}
\setcounter{table}{0}
\renewcommand{\thetable}{A.\arabic{table}}
\setcounter{figure}{0}
\renewcommand{\thefigure}{A.\arabic{figure}}
\begin{appendix}
\section{Instructions}

English-language instructions are available here: \url{https://cdn.mg.sb/upload/s/instructions-heating-law-complete.pdf}

\section{Motives}
\label{app:motives}

    \emph{The order of motives was randomized between subjects, but kept constant within subjects.}

\begin{table}[H]
\centering
\begin{tabular}{@{}ll@{}}
\toprule
Abbreviation & Motive                                                     \\ \midrule
System works well    & My current heating system still works well.                \\
Uncertainty  & The government's actions are too unpredictable.            \\
             & I don't believe the Law will stay like this.               \\
Effort    & Switching is too much effort.                              \\
Reliability     & I believe that my current heating system is more reliable. \\
Not alone    & I cannot decide about the heating system by myself;        \\
             & the other deciders are against switching.                  \\
Expenses    & Switching is too expensive.                                \\
Pressure     & I feel that I am being bullied/railroaded by the Law.      \\
Other        &                                                            \\ \bottomrule
\end{tabular}
\caption{Motives presented to fossil users}\label{motives4fossil}
\end{table}

\begin{table}[H]
\centering
\begin{tabular}{@{}ll@{}}
\toprule
Abbreviation & Motive                                             \\ \midrule
Common good  & I want to contribute to climate protection.        \\
Independence & I want us to become independent of fossil imports. \\
Expenses    & Switching is worth it financially.                 \\
Pressure     & I was bullied/railroaded into switching.           \\
Other        &                                                    \\ \bottomrule
\end{tabular}
\caption{Motives presented to non-fossil users}\label{motives4nonfossil}

\end{table}

\section{Policy goals}
\label{app:policy}

    \emph{The order of policy goals was randomized between subjects, but kept constant within subjects.}

\begin{table}[H]
\centering
\begin{tabular}{@{}ll@{}}
\toprule
Abbreviation & Goal                                             \\ \midrule
Growth       & Economic growth                                  \\
Inequality   & Reducing inequality                              \\
Freedom      & Freedom                                          \\
Environment* & Preserve the environment, the climate and nature \\
Traditions   & Preserve traditions                              \\
Security     & Peace and security                               \\
Other        &                                                  \\ \bottomrule
\end{tabular}
\end{table}

*This goal was used to construct the variable “Pro-Environmental Priority,” see Section \ref{sec:measures}.
	
\section{Additional tables and figures}

\begin{table}
\centering
\begin{tabular}{@{}p{0.15\textwidth}p{0.8\textwidth}@{}}
\toprule
Question ID & Question text                                                                                                                                                                                                       \\ \midrule
Q1          & Heat pumps installed in a single apartment or in a privately used single-family home with only one apartment must also be inspected annually by a chimney sweep or another qualified person, just like gas heaters. \\
Q2          & A new gas heater may still be installed today in a house built in 1960.                                                                                                                                             \\
Q3          & A new gas heater may still be installed in 2029 in a house built in 1960.                                                                                                                                           \\
Q4          & By mid-2028 at the latest, only heating systems that obtain at least 65\% of their energy from renewable sources may be installed.                                                                                  \\
Q5          & A gas heater installed on January 1, 2000, in a rented apartment may still be used in the year 2032.                                                                                                                \\
Q6          & If you operate a gas heater beyond its permissible usage duration, fines in the five-figure range may be incurred.                                                                                                  \\
Q7          & A permissible gas heater installed today may still be operated in the year 2040.                                                                                                                                    \\
Q8          & Homeowners will receive a subsidy of at least 30\%, but no more than €10,000, for the installation of a new heat pump in 2024.                                                                                      \\
Q9          & In new buildings, the installation of gas and oil heaters is permitted without restrictions until 2028.                                                                                                             \\
Q10         & The use of conventional heating systems powered by fossil fuels is allowed indefinitely.                                                                                                                            \\
Q11         & A heat pump uses electricity to generate heat.                                                                                                                                                                      \\
Q12         & When heating with a heat pump, the cost of heating is determined by the price of electricity.                                                                                                                       \\ \bottomrule
\end{tabular}

\caption{Questions used in the quiz}
\label{tquestions}
\end{table}

\begin{table}
\begin{center}
\sisetup{parse-numbers=false, table-text-alignment=center}
\begin{tabular}{l S[table-format=4.6] S[table-format=4.6]}
\toprule
 & {Knowledge Score} & {Knowledge Score} \\
\midrule
\emph{Intercept}               & 6.184^{***} & 4.680^{***}  \\
                               & (0.083)     & (0.239)      \\
\emph{Treatment}: Info Full    & 0.919^{***} & 0.924^{***}  \\
                               & (0.122)     & (0.119)      \\
\emph{Treatment}: Info Strict  & 0.736^{***} & 0.723^{***}  \\
                               & (0.121)     & (0.118)      \\
\emph{Treatment}: Info Lenient & -0.032      & -0.033       \\
                               & (0.118)     & (0.116)      \\
Female                         &             & -0.544^{***} \\
                               &             & (0.090)      \\
Age                            &             & 0.023^{***}  \\
                               &             & (0.003)      \\
High education                 &             & 0.124        \\
                               &             & (0.091)      \\
High income                    &             & 0.379^{***}  \\
                               &             & (0.088)      \\
Household size                 &             & 0.000        \\
                               &             & (0.000)      \\
Landlord                       &             & 0.042        \\
                               &             & (0.097)      \\
Duplex house                   &             & 0.724^{***}  \\
                               &             & (0.153)      \\
Single-family house            &             & 0.586^{***}  \\
                               &             & (0.116)      \\
Other home                     &             & 1.352^{*}    \\
                               &             & (0.714)      \\
Townhouse                      &             & 0.550^{***}  \\
                               &             & (0.162)      \\
Sole decider                   &             & -0.100       \\
                               &             & (0.101)      \\
\midrule
R$^2$                          & 0.036       & 0.089        \\
Adj. R$^2$                     & 0.035       & 0.085        \\
Num. obs.                      &{2668}       &{2668}        \\
\bottomrule
\multicolumn{3}{l}{\scriptsize{$^{***}p<0.01$; $^{**}p<0.05$; $^{*}p<0.1$}}
\end{tabular}
\caption{OLS regressions of knowledge on treatments and demographic variables (manipulation check)\\{\footnotesize \emph{Knowledge Score} ranges from 0 to 12. HC3 standard errors in parentheses.}}
\label{regknow}
\end{center}
\end{table}

\begin{table}
\begin{center}
\sisetup{parse-numbers=false, table-text-alignment=center}
\begin{tabular}{l S[table-format=4.6] S[table-format=4.6] S[table-format=4.6] S[table-format=4.6]}
\toprule
 & {Plan fossil usage} & {Law sensible} & {Give to climate org} & {Give to pro-liberty org} \\
\midrule
\emph{Intercept}    & 0.465^{***}  & 2.631^{***}  & 16.028^{***} & 21.351^{***} \\
                    & (0.054)      & (0.145)      & (2.659)      & (2.351)      \\
Knowledge Score     & -0.002       & 0.007        & 0.323        & -0.047       \\
                    & (0.004)      & (0.011)      & (0.208)      & (0.208)      \\
Female              & 0.032        & 0.029        & 1.605        & 0.740        \\
                    & (0.020)      & (0.054)      & (0.994)      & (0.907)      \\
Age                 & 0.004^{***}  & -0.008^{***} & 0.045        & -0.021       \\
                    & (0.001)      & (0.002)      & (0.039)      & (0.034)      \\
High education      & -0.031       & 0.445^{***}  & 5.294^{***}  & -0.790       \\
                    & (0.020)      & (0.055)      & (1.007)      & (0.925)      \\
High income         & -0.067^{***} & 0.104^{*}    & 0.950        & -0.193       \\
                    & (0.020)      & (0.054)      & (0.987)      & (0.893)      \\
Household size      & -0.000       & 0.000        & 0.001        & -0.000       \\
                    & (0.000)      & (0.001)      & (0.002)      & (0.006)      \\
Landlord            & -0.002       & 0.270^{***}  & 2.597^{**}   & 3.797^{***}  \\
                    & (0.022)      & (0.061)      & (1.087)      & (0.995)      \\
Duplex house        & 0.054        & 0.024        & 1.856        & 3.042^{**}   \\
                    & (0.034)      & (0.093)      & (1.625)      & (1.548)      \\
Single-family house & 0.025        & -0.049       & 0.419        & 0.146        \\
                    & (0.026)      & (0.072)      & (1.260)      & (1.128)      \\
Other home          & 0.173        & -0.219       & 0.782        & 3.549        \\
                    & (0.124)      & (0.388)      & (5.592)      & (7.543)      \\
Townhouse           & 0.128^{***}  & -0.068       & 2.759        & 1.786        \\
                    & (0.035)      & (0.100)      & (1.835)      & (1.567)      \\
Sole decider        & 0.010        & 0.064        & -2.118^{*}   & -0.469       \\
                    & (0.022)      & (0.060)      & (1.137)      & (1.044)      \\
\midrule
R$^2$               & 0.028        & 0.062        & 0.021        & 0.009        \\
Adj. R$^2$          & 0.024        & 0.057        & 0.017        & 0.004        \\
Num. obs.           &{2668}        &{2668}        &{2668}        &{2668}        \\
\bottomrule
\multicolumn{5}{l}{\scriptsize{$^{***}p<0.01$; $^{**}p<0.05$; $^{*}p<0.1$}}
\end{tabular}
\caption{OLS regressions of behavior and attitudes on knowledge and demographic variables\\{\footnotesize \emph{Plan fossil usage} is a binary variable that merges decisions from landlords and proprietors. \emph{Law sensible} is a Likert scale that ranges from 1 to 5. \emph{Give to climate org} is a euro amount (€0--€100) donated to the climate charity in the dictator game. Similarly, \emph{Give to pro-liberty org} is a euro amount (€0--€100) donated to the pro-liberty charity. HC3 standard errors in parentheses.}}
\label{regt1k}
\end{center}
\end{table}

\begin{figure}[H]
\centering
\includegraphics[width=0.99\textwidth]{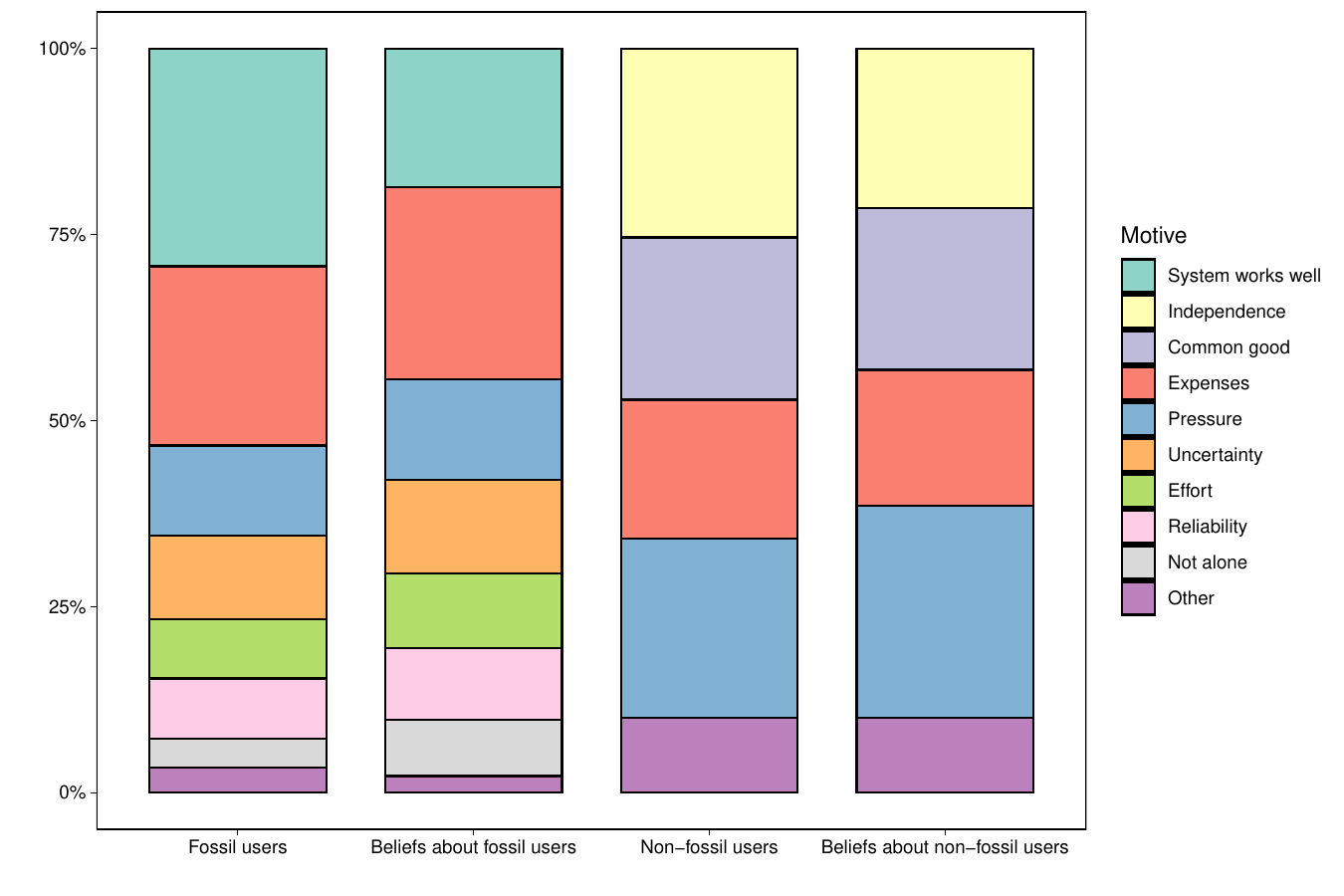}
    \caption{Motives for (not) switching to a new heating system\\{\footnotesize This Figure shows average motives and average beliefs about others' motives. Once again, we merge data from proprietors and landlords.}}
\label{fig:combined_motives}
\end{figure}

\begin{figure}[H]
\centering
\includegraphics[width=0.99\textwidth]{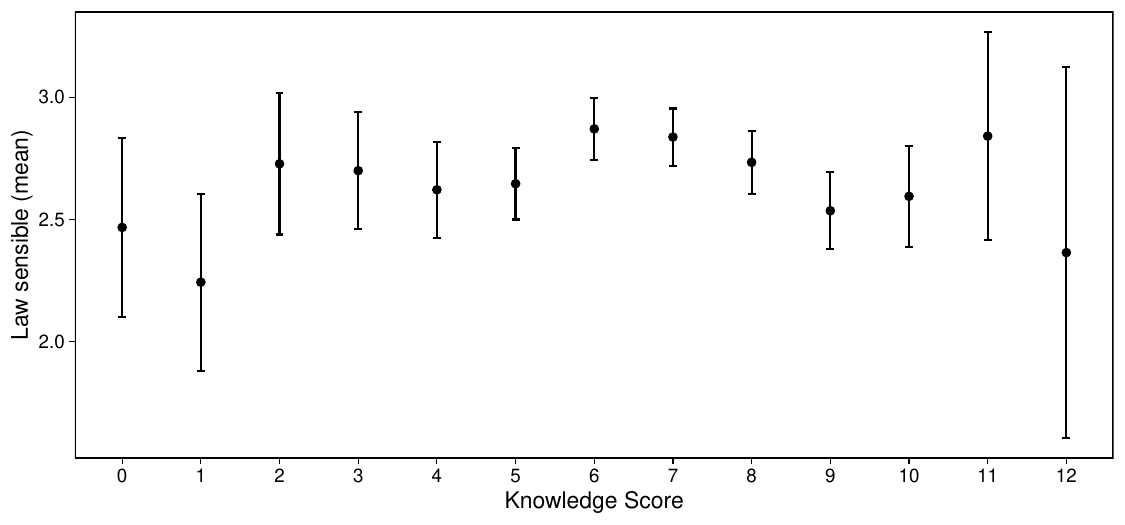}
    \caption{Relationship between knowledge and assessment of the Law}
\label{fig:knowfine}
\end{figure}

\begin{figure}[H]
\centering
\includegraphics[width=0.99\textwidth]{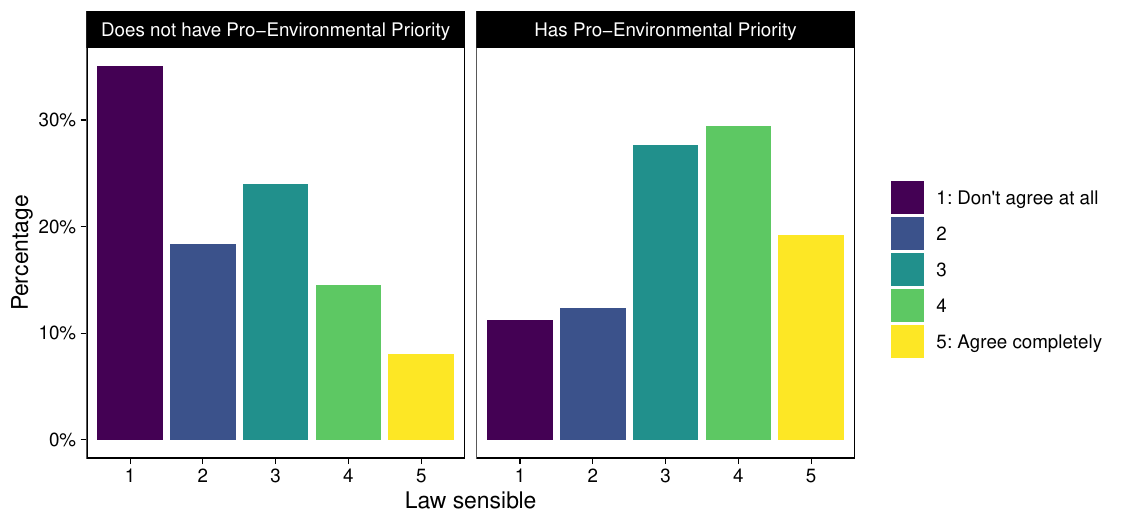}
    \caption{Relationship between Pro-Environmental Priority and assessment of the Law}
\label{fig:priofine}
\end{figure}

\begin{table}
\begin{center}
\begin{tabular}[t]{lccccc}
\toprule
Variable & Min. & Max. & Median & Mean & SD\\
\midrule
\textit{Fossil users}:\\
System works well & 0 & 100 & 20 & 29.26 & 29.70\\
Pressure & 0 & 100 & 5 & 12.09 & 18.08\\
Uncertainty & 0 & 100 & 5 & 11.25 & 15.59\\
Expenses & 0 & 100 & 20 & 24.09 & 24.64\\
Effort & 0 & 100 & 4 & 7.96 & 12.02\\
Reliability & 0 & 100 & 0 & 8.12 & 14.24\\
Not alone & 0 & 100 & 0 & 3.88 & 11.99\\
{Other} & 0 & 100 & 0 & 3.35 & 14.54\\
\addlinespace
\textit{Beliefs about fossil users}:\\
System works well & 0 & 100 & 15 & 18.64 & 16.14\\
Pressure & 0 & 100 & 10 & 13.49 & 14.49\\
Uncertainty & 0 & 100 & 10 & 12.58 & 13.51\\
Expenses & 0 & 100 & 20 & 25.84 & 20.68\\
Effort & 0 & 100 & 10 & 10.05 & 10.61\\
Reliability & 0 & 100 & 10 & 9.66 & 11.24\\
Not alone & 0 & 56 & 5 & 7.51 & 9.05\\
{Other} & 0 & 100 & 0 & 2.23 & 8.51\\
\addlinespace
\textit{Non-fossil users}:\\
Common good & 0 & 100 & 20 & 21.80 & 21.01\\
Independence & 0 & 100 & 20 & 25.40 & 22.66\\
Expenses & 0 & 100 & 10 & 18.69 & 22.55\\
Pressure & 0 & 100 & 10 & 24.06 & 30.36\\
{Other} & 0 & 100 & 0 & 10.05 & 23.90\\
\addlinespace
\textit{Beliefs about non-fossil users}:\\
Common good & 0 & 100 & 20 & 21.71 & 15.57\\
Independence & 0 & 100 & 20 & 21.47 & 16.46\\
Expenses & 0 & 100 & 15 & 18.29 & 17.35\\
Pressure & 0 & 100 & 20 & 28.44 & 24.81\\
{Other} & 0 & 100 & 1 & 10.09 & 16.43\\
\bottomrule
\end{tabular}

\caption{Summary statistics about motives}
\label{mot1tab}
\end{center}
\end{table}

\begin{table}
\begin{center}
\sisetup{parse-numbers=false, table-text-alignment=center}
\resizebox{\textwidth}{!}{%
\begin{tabular}{l S[table-format=3.6] S[table-format=3.6] S[table-format=3.6] S[table-format=3.6]}
\toprule
 & {Plan fossil usage} & {Law sensible} & {Give to climate org} & {Give to pro-liberty org} \\
\midrule
\emph{Intercept}    & 0.454^{***} & 3.284^{***}  & 24.094^{***} & 25.551^{***} \\
                    & (0.128)     & (0.344)      & (5.939)      & (5.059)      \\
Knowledge Score     & -0.002      & -0.025       & -0.153       & 0.168        \\
                    & (0.009)     & (0.024)      & (0.440)      & (0.432)      \\
Female              & 0.004       & 0.087        & 1.209        & -0.494       \\
                    & (0.041)     & (0.105)      & (1.999)      & (1.853)      \\
Age                 & 0.005^{***} & -0.018^{***} & -0.074       & -0.081       \\
                    & (0.002)     & (0.004)      & (0.081)      & (0.071)      \\
High education      & -0.077^{*}  & 0.369^{***}  & 3.354^{*}    & -1.691       \\
                    & (0.040)     & (0.105)      & (2.031)      & (1.898)      \\
High income         & -0.008      & -0.112       & -1.370       & -0.028       \\
                    & (0.041)     & (0.109)      & (2.023)      & (1.864)      \\
Household size      & -0.000      & 0.000        & 0.001        & -0.000       \\
                    & (0.017)     & (0.053)      & (0.432)      & (0.019)      \\
Landlord            & -0.040      & 0.303^{**}   & 6.267^{***}  & 6.560^{***}  \\
                    & (0.046)     & (0.125)      & (2.187)      & (2.060)      \\
Duplex house        & -0.001      & 0.122        & 5.908^{*}    & 1.425        \\
                    & (0.070)     & (0.174)      & (3.214)      & (3.166)      \\
Single-family house & -0.052      & 0.115        & 4.099^{*}    & 0.225        \\
                    & (0.056)     & (0.144)      & (2.393)      & (2.498)      \\
Other home          & 0.154       & -0.333       & -6.109       & -13.767^{*}  \\
                    & (0.300)     & (0.683)      & (11.492)     & (7.639)      \\
Townhouse           & 0.070       & -0.098       & 2.151        & 0.196        \\
                    & (0.071)     & (0.196)      & (3.340)      & (3.134)      \\
Sole decider        & 0.017       & 0.155        & -3.532       & -1.964       \\
                    & (0.044)     & (0.112)      & (2.147)      & (2.057)      \\
\midrule
R$^2$               & 0.046       & 0.093        & 0.036        & 0.023        \\
Adj. R$^2$          & 0.029       & 0.076        & 0.019        & 0.005        \\
Num. obs.           &{674}        &{674}         &{674}         &{674}         \\
\bottomrule
\multicolumn{5}{l}{\scriptsize{$^{***}p<0.01$; $^{**}p<0.05$; $^{*}p<0.1$}}
\end{tabular}}
\caption{OLS regressions of behavior and attitudes on treatments and demographic variables in the Baseline treatment\\{\footnotesize \emph{Plan fossil usage} is a binary variable that merges decisions from landlords and proprietors. \emph{Law sensible} is a Likert scale that ranges from 1 to 5. \emph{Give to climate org} is a euro amount (€0--€100) donated to the climate charity in the dictator game. Similarly, \emph{Give to pro-liberty org} is a euro amount (€0--€100) donated to the pro-liberty charity. HC3 standard errors in parentheses.}}
\label{regt1base}
\end{center}
\end{table}

\begin{table}
\begin{center}
\sisetup{parse-numbers=false, table-text-alignment=center}
\resizebox{\textwidth}{!}{%
\begin{tabular}{l S[table-format=4.6] S[table-format=4.6] S[table-format=4.6] S[table-format=4.6] S[table-format=4.6] S[table-format=4.6] S[table-format=4.6]}
\toprule
 & {Growth} & {Inequality} & {Freedom} & {Environment} & {Traditions} & {Security} & {Pro-Environmental Priority} \\
\midrule
\emph{Intercept}               & 18.878^{***} & 17.239^{***} & 18.832^{***} & 12.733^{***} & 11.499^{***} & 14.515^{***} & 0.290^{***} \\
                               & (1.625)      & (1.382)      & (1.296)      & (1.608)      & (0.879)      & (1.539)      & (0.051)     \\
\emph{Treatment}: Info Full    & 0.759        & 0.261        & -0.316       & -1.187       & -0.045       & 0.525        & -0.048^{*}  \\
                               & (0.771)      & (0.651)      & (0.655)      & (0.792)      & (0.451)      & (0.836)      & (0.025)     \\
\emph{Treatment}: Info Strict  & -0.195       & -0.571       & 0.765        & -0.670       & 0.246        & 0.415        & -0.013      \\
                               & (0.754)      & (0.592)      & (0.705)      & (0.823)      & (0.476)      & (0.848)      & (0.026)     \\
\emph{Treatment}: Info Lenient & 0.000        & 0.196        & -0.226       & -0.129       & 0.743        & -0.651       & -0.003      \\
                               & (0.762)      & (0.624)      & (0.671)      & (0.846)      & (0.503)      & (0.839)      & (0.026)     \\
Female                         & -1.681^{***} & 0.288        & -2.107^{***} & 1.003^{*}    & -0.782^{**}  & 3.397^{***}  & 0.012       \\
                               & (0.580)      & (0.466)      & (0.507)      & (0.589)      & (0.347)      & (0.644)      & (0.019)     \\
Age                            & -0.020       & -0.087^{***} & -0.034^{*}   & 0.101^{***}  & -0.090^{***} & 0.192^{***}  & 0.000       \\
                               & (0.024)      & (0.018)      & (0.019)      & (0.025)      & (0.013)      & (0.023)      & (0.001)     \\
High education                 & 0.471        & -0.127       & -0.504       & 2.582^{***}  & -1.936^{***} & 0.192        & 0.064^{***} \\
                               & (0.587)      & (0.504)      & (0.554)      & (0.637)      & (0.369)      & (0.660)      & (0.019)     \\
High income                    & 1.356^{**}   & -0.302       & -0.129       & -0.086       & 0.069        & -0.578       & 0.003       \\
                               & (0.559)      & (0.480)      & (0.515)      & (0.596)      & (0.347)      & (0.616)      & (0.019)     \\
Household size                 & 0.000        & -0.000       & -0.000       & 0.000        & -0.000       & 0.000        & -0.000      \\
                               & (0.004)      & (0.005)      & (0.002)      & (0.002)      & (0.000)      & (0.005)      & (0.000)     \\
Landlord                       & -1.279^{**}  & -0.639       & 0.461        & 0.625        & 1.203^{***}  & -1.215^{*}   & 0.025       \\
                               & (0.623)      & (0.487)      & (0.563)      & (0.701)      & (0.399)      & (0.669)      & (0.021)     \\
Landlord                       & -1.279^{**}  & -0.639       & 0.461        & 0.625        & 1.203^{***}  & -1.215^{*}   & 0.025       \\
                               & (0.623)      & (0.487)      & (0.563)      & (0.701)      & (0.399)      & (0.669)      & (0.021)     \\
Duplex house                   & 2.269^{**}   & -1.831^{**}  & -2.108^{***} & 0.977        & 0.940        & -0.536       & -0.018      \\
                               & (1.055)      & (0.797)      & (0.799)      & (1.080)      & (0.585)      & (1.009)      & (0.032)     \\
Single-family house            & -0.023       & -0.893       & -0.280       & -0.496       & 1.167^{***}  & 0.381        & -0.005      \\
                               & (0.739)      & (0.647)      & (0.694)      & (0.785)      & (0.432)      & (0.814)      & (0.025)     \\
Other home                     & -4.873^{**}  & 2.228        & 4.073        & 1.792        & 2.965        & -5.030^{*}   & -0.000      \\
                               & (2.334)      & (4.174)      & (3.530)      & (4.716)      & (2.323)      & (2.963)      & (0.141)     \\
Townhouse                      & -0.234       & -0.503       & -1.415^{*}   & -0.173       & 0.822        & 1.402        & 0.026       \\
                               & (0.961)      & (0.869)      & (0.858)      & (1.114)      & (0.604)      & (1.195)      & (0.036)     \\
Sole decider                   & -0.488       & -0.199       & 0.915^{*}    & -0.918       & 1.370^{***}  & -1.033       & -0.016      \\
                               & (0.642)      & (0.538)      & (0.544)      & (0.671)      & (0.358)      & (0.739)      & (0.022)     \\
\midrule
R$^2$                          & 0.014        & 0.013        & 0.017        & 0.016        & 0.042        & 0.049        & 0.008       \\
Adj. R$^2$                     & 0.008        & 0.008        & 0.012        & 0.011        & 0.037        & 0.044        & 0.003       \\
Num. obs.                      &{2668}        &{2668}        &{2668}        &{2668}        &{2668}        &{2668}        &{2668}       \\
\bottomrule
\multicolumn{8}{l}{\scriptsize{$^{***}p<0.01$; $^{**}p<0.05$; $^{*}p<0.1$}}
\end{tabular}}
\caption{OLS regressions of policy goals on treatments and demographic variables\\{\footnotesize See Section \ref{app:policy} in the Appendix and Section \ref{sec:measures} in the main text for a description of the policy goals and their elicitation. HC3 standard errors in parentheses.}}
\label{regtgoaltreat}
\end{center}
\end{table}

\begin{table}
\begin{center}
\sisetup{parse-numbers=false, table-text-alignment=center}
\resizebox{\textwidth}{!}{%
\begin{tabular}{l S[table-format=4.6] S[table-format=4.6] S[table-format=4.6] S[table-format=4.6]}
\toprule
 & {Plan fossil usage} & {Law sensible} & {Give to climate org} & {Give to pro-liberty org} \\
\midrule
\emph{Intercept}               & 0.495^{***} & 3.064^{***}  & 14.208^{***} & 18.946^{***} \\
                               & (0.118)     & (0.309)      & (5.113)      & (4.858)      \\
\emph{Treatment}: Info Full    & 0.009       & -0.010       & 0.672        & -1.021       \\
                               & (0.026)     & (0.066)      & (1.244)      & (1.192)      \\
\emph{Treatment}: Info Strict  & -0.004      & -0.041       & 1.568        & -1.319       \\
                               & (0.027)     & (0.069)      & (1.291)      & (1.192)      \\
\emph{Treatment}: Info Lenient & 0.011       & -0.024       & 0.560        & -1.587       \\
                               & (0.026)     & (0.068)      & (1.244)      & (1.175)      \\
\emph{Goal}: Growth            & 0.001       & -0.013^{***} & -0.156^{***} & -0.033       \\
                               & (0.001)     & (0.003)      & (0.052)      & (0.053)      \\
\emph{Goal}: Inequality        & -0.000      & 0.004        & 0.135^{**}   & 0.027        \\
                               & (0.001)     & (0.004)      & (0.062)      & (0.060)      \\
\emph{Goal}: Freedom           & 0.001       & -0.011^{***} & -0.034       & 0.119^{**}   \\
                               & (0.001)     & (0.003)      & (0.055)      & (0.058)      \\
\emph{Goal}: Environment       & -0.003^{**} & 0.021^{***}  & 0.388^{***}  & -0.079       \\
                               & (0.001)     & (0.004)      & (0.061)      & (0.050)      \\
\emph{Goal}: Traditions        & -0.002      & -0.012^{***} & -0.123^{*}   & 0.204^{***}  \\
                               & (0.002)     & (0.004)      & (0.064)      & (0.077)      \\
\emph{Goal}: Security          & -0.000      & -0.007^{**}  & 0.054        & -0.023       \\
                               & (0.001)     & (0.003)      & (0.051)      & (0.049)      \\
\midrule
Demographic controls           & {Yes}       & {Yes}        & {Yes}        & {Yes}        \\
R$^2$                          & 0.038       & 0.186        & 0.107        & 0.028        \\
Adj. R$^2$                     & 0.031       & 0.180        & 0.100        & 0.021        \\
Num. obs.                      &{2668}       &{2668}        &{2668}        &{2668}        \\
\bottomrule
\multicolumn{5}{l}{\scriptsize{$^{***}p<0.01$; $^{**}p<0.05$; $^{*}p<0.1$}}
\end{tabular}}
\caption{OLS regressions of behavior and attitudes on all policy goals except for Other (see Section \ref{app:policy} in the Appendix)\\{\footnotesize \emph{Plan fossil usage} is a binary variable that merges decisions from landlords and proprietors. \emph{Law sensible} is a Likert scale that ranges from 1 to 5. \emph{Give to climate org} is a euro amount (€0--€100) donated to the climate charity in the dictator game. Similarly, \emph{Give to pro-liberty org} is a euro amount (€0--€100) donated to the pro-liberty charity. Demographic controls: Female, age, high education, high income, household size, landlord, duplex house, single-family house, other home, townhouse, sole decider. HC3 standard errors in parentheses.}}
\label{regt2}
\end{center}
\end{table}

\begin{table}
\begin{center}
\sisetup{parse-numbers=false, table-text-alignment=center}
\begin{tabular}{l S[table-format=4.6] S[table-format=4.6] S[table-format=4.6] S[table-format=4.6]}
\toprule
 & {Plan fossil usage} & {Law sensible} & {Give to climate org} & {Give to pro-liberty org} \\
\midrule
\emph{Intercept}           & 0.472^{***}  & 2.313^{***}  & 12.318^{***} & 22.664^{***} \\
                           & (0.057)      & (0.146)      & (2.783)      & (2.498)      \\
Q1 correct                 & -0.110^{***} & -0.055       & 1.425        & -2.850^{***} \\
                           & (0.020)      & (0.054)      & (1.040)      & (0.921)      \\
Q2 correct                 & 0.039^{*}    & 0.010        & 1.770^{*}    & 0.950        \\
                           & (0.020)      & (0.052)      & (0.987)      & (0.895)      \\
Q3 correct                 & 0.027        & 0.031        & 1.102        & 0.865        \\
                           & (0.020)      & (0.052)      & (0.973)      & (0.875)      \\
Q4 correct                 & -0.012       & 0.104^{*}    & 1.867        & -0.320       \\
                           & (0.023)      & (0.061)      & (1.146)      & (1.059)      \\
Q5 correct                 & 0.016        & 0.168^{**}   & -0.244       & 0.161        \\
                           & (0.027)      & (0.072)      & (1.237)      & (1.107)      \\
Q6 correct                 & -0.037^{*}   & -0.096^{*}   & -0.955       & 0.357        \\
                           & (0.021)      & (0.054)      & (0.985)      & (0.937)      \\
Q7 correct                 & 0.054^{***}  & 0.044        & -0.129       & -0.471       \\
                           & (0.020)      & (0.054)      & (0.982)      & (0.911)      \\
Q8 correct                 & -0.043^{**}  & 0.246^{***}  & 1.841^{*}    & 0.053        \\
                           & (0.019)      & (0.051)      & (0.960)      & (0.876)      \\
Q9 correct                 & 0.052^{***}  & -0.090^{*}   & -0.167       & 0.934        \\
                           & (0.019)      & (0.051)      & (0.977)      & (0.879)      \\
Q10 correct                & 0.002        & -0.139^{***} & -0.206       & -0.338       \\
                           & (0.020)      & (0.051)      & (0.977)      & (0.887)      \\
Q11 correct                & -0.038       & 0.091        & 0.049        & 1.207        \\
                           & (0.025)      & (0.067)      & (1.234)      & (1.090)      \\
Q12 correct                & 0.016        & -0.087       & -1.953       & -1.176       \\
                           & (0.026)      & (0.066)      & (1.210)      & (1.076)      \\
Pro-Environmental Priority & -0.092^{***} & 0.861^{***}  & 11.323^{***} & -3.156^{***} \\
                           & (0.020)      & (0.052)      & (1.033)      & (0.839)      \\
\midrule
Demographic controls       & {Yes}        & {Yes}        & {Yes}        & {Yes}        \\
R$^2$                      & 0.057        & 0.167        & 0.075        & 0.019        \\
Adj. R$^2$                 & 0.048        & 0.159        & 0.066        & 0.009        \\
Num. obs.                  &{2668}        &{2668}        &{2668}        &{2668}        \\
\bottomrule
\multicolumn{5}{l}{\scriptsize{$^{***}p<0.01$; $^{**}p<0.05$; $^{*}p<0.1$}}
\end{tabular}
\caption{OLS regressions of behavior and attitudes on individual question responses and all policy goals except for Other (see Section \ref{app:policy} in the Appendix)\\{\footnotesize \emph{Plan fossil usage} is a binary variable that merges decisions from landlords and proprietors. \emph{Law sensible} is a Likert scale that ranges from 1 to 5. \emph{Give to climate org} is a euro amount (€0--€100) donated to the climate charity in the dictator game. Similarly, \emph{Give to pro-liberty org} is a euro amount (€0--€100) donated to the pro-liberty charity. Demographic controls: Female, age, high education, high income, household size, landlord, duplex house, single-family house, other home, townhouse, sole decider. HC3 standard errors in parentheses.}}
\label{regt9}
\end{center}
\end{table}

\end{appendix}
\end{document}